\documentclass[12pt]{article} 
\usepackage{amsmath}
\usepackage{amsfonts}
\usepackage{amssymb}
\usepackage{graphicx}
\usepackage{color}
\usepackage{wrapfig}
\usepackage{pifont}
\usepackage{epstopdf}
\usepackage{booktabs}

\newtheorem{definition}{Definition}[section]

\newtheorem{theorem}{Theorem}[section]
\newtheorem{corollary}{Corollary}[section]
\newtheorem{proposition}{Proposition}[section]

\newcommand{\cD}{\mathcal{D}}

\newenvironment{proofproof}{{\noindent\bf Proof\ }}{\hfill{$\Box$}\vspace{0.1in}}

\title{
 Locating a Phylogenetic Tree in A Reticulation-Visible Network in Quadratic Time }
\author{Andreas D.M. Gunawan\thanks{Department of Mathematics, National University of Singapore, Singapore 119076, Singapore.}, Bhaskar DasGupta\thanks{Department of Computer Science, University of Illinois at Chicago, Chicago, IL 60607, USA.}, Louxin Zhang\thanks{To whom correspondence should be addressed. Department of Mathematics, National University of Singapore, Singapore 119076, Singapore. E-mail: matzlx@nus.edu.sg. }}
\date{}

\begin{document}
\maketitle

\begin{abstract}
	In phylogenetics, phylogenetic trees are rooted binary trees,  whereas phylogenetic networks are rooted arbitrary acyclic digraphs.   Edges are directed away from the root and leaves are uniquely labeled with taxa in phylogenetic networks. For the purpose of validating evolutionary models, biologists check whether or not  a  phylogenetic tree is contained in a  phylogenetic network on the same taxa. This tree containment problem is known to be NP-complete. A phylogenetic network is reticulation-visible if every reticulation node separates the root of the network  from some leaves. We answer an open problem by proving that the problem is solvable in quadratic time for reticulation-visible networks. The key tool used in our answer is a powerful decomposition theorem. It also allows us to design a linear-time algorithm for the cluster containment problem for networks of this type and to prove that every galled network with $n$ leaves has $2(n-1)$ reticulation nodes at most.
\end{abstract}

\section{Introduction}
	How life came to existence and evolved  has been a key scientific question in the past hundred or so years.  
Traditionally, a phylogenetic tree has been used to model the evolutionary history of species, in which  an internal  node  represents a \emph{speciation event} and the leaves represent the extant species under study. These evolutionary trees are often reconstructed from the gene or protein sequences sampled from the extant  species. Since genomic studies have demonstrated that genetic material is often transfered between organisms in a non-reproductive manner \cite{Chan_13_PNAS,Treangen_11_PLOSGenentics}, it has been commonly accepted that phylogenetic networks are more suitable than phylogenetic trees  for modeling horizontal gene transfer, introgression, recombination and hybridization events in genome evolution \cite{Dagan_08_PNAS,Doolittle,Gusfield_14_Book,Moret_04_TCBB,Nakhleh_13_TREE}. Mathematically, a phylogenetic network is a rooted acyclic digraph with uniquely labeled leaves. 
The algorithmic and combinatorial aspects of  networks have been intensively studied over the past two decades ({\it e.g.}, see \cite{Gusfield_14_Book,Huson_book,Wang_01_JCB}).
	
	In phylogenetics, an important  issue is checking the ``consistency'' of two evolutionary models. A somewhat simpler (but nonetheless very important) version of  this issue asks whether a given network is consistent with an existing tree model or not.  This has motivated researchers to study the problem of determining whether a tree is displayed by a network or not, which is called the tree containment problem (TCP). The cluster containment problem (CCP) is related algorithmic problem that asks whether or not a subset of taxa is a cluster in a tree displayed by a network.  Both the TCP and CCP have also been investigated in the development of network metrics \cite{Cardona_09_TCBB,Kanj_08_TCS} 

Both the TCP and CCP are NP-complete \cite{Kanj_08_TCS},  even on a very restricted class of  networks \cite{van_Iersel_2010_IPL}.  
van Iersel \textit{et al}. have posed an open problem as to whether or not the TCP is solvable in polynomial time for reticulation-visible networks
\cite{Gunawan_2015, Huson_book,van_Iersel_2010_IPL}. The visibility property was originally introduced to capture an important feature of galled networks \cite{Huson_Recomb}. A  network is reticulation-visible if every reticulation node separates the network root from some leaves.  Real network models are likely to be reticulation-visible (see  \cite{Marcussen_SysBiol_12} for example). Although much effort has been devoted to the study of the TCP, it has been shown to be solvable in polynomial-time only for a couple of very restricted subclasses of reticulation-visible networks \cite{Philippe_2015,van_Iersel_2010_IPL}. Other studies related to the TCP include \cite{Semple}, \cite{Linz_Count} and \cite{Wilson_2011}.
 
 In this paper,  we make three  contributions. We give an affirmative answer to the open problem by presenting a 
quadratic time algorithm for the TCP for binary reticulation-visible networks. Additionally, we  present a linear-time algorithm for the CCP for arbitrary reticulation-visible networks.  Our algorithms rely on an important  decomposition theorem
(Theorems~\ref{main-theorem} and \ref{main-theorem2}), which is proven in Section~\ref{sec4}.  Empowered by this, we also prove that an arbitrary galled network with $n$ leaves has 
$2(n-1)$ reticulation nodes at most. 

 The rest of the paper is organized as follows. 
 Section~\ref{sec2} introduces the basic concepts and notation. 
 In Section~\ref{sec4}, we present a decomposition theorem (Theorem ~\ref{Decomp_Thm}) that reveals an important structural property of reticulation-visible networks, based on which the two main theorems  (Theorem ~\ref{Decomp_Thm}) are  proven in Sections~\ref{sec5} and \ref{sec6}, respectively. 
 Finally, we conclude with a couple of remarks in Section~\ref{sec8}.
	
\section{Basic Concepts and Notation}
\label{sec2}

\subsection{Phylogenetic networks}

In phylogenetics, \emph{networks} are rooted acyclic digraphs in which a unique node (the \emph{root}) exists such that there is a directed path from it to \emph{every} other node and the nodes of indegree one and outdegree zero (the {\it leaves}) are \emph{uniquely} labeled. The leaf labels represent  bio-molecular sequences, extant organisms or species under study. 

In a network,  the root has indegree zero and outdegree greater than one and  each of the other nodes has  either indegree one or  outdegree one exclusively.
 A node is called a {\it reticulation} (node) if its indegree is strictly greater than one and its outdegree is precisely one.  A reticulation node is called a {\it bicombination} if it has indegree two. Reticulation nodes represent reticulation events occurring in evolution.  
 Non-reticulation nodes are called {\it tree} nodes, which include the root and leaves.

 For convenience in describing the algorithms and proofs, we add an \emph{open} incoming edge to the root (Figure~\ref{example1}).  A network is called a {\it bicombining} network if
every reticulation node is of degree three ({\it i.e.}, indegree two and outdegree one). 
A network is called  \emph{binary}  if the root is of degree two,  its leaves are of degree one,  and  all other nodes are of degree three. 

Let $N$ be a network.
For two nodes $u, v$ in $N$,  $u$ is a \emph{parent} of $v$ (alternatively,  $v$ is a \emph{child} of $u$)  if $(u, v)$ is a directed edge in $N$; 
$u$ is an \emph{ancestor} of  $v$   (alternatively,  $v$ is a \emph{descendant} of $u$)  if there is a directed path from $u$ to $v$.  When $u$ is an ancestor of $v$,  we also say  $v$ is {\it below} $u$ and $u$ is {\it above} $v$.

%

Let $N$ be a network. We use the following notation:\vspace{-0.5em}
\begin{itemize}
\item $\rho(N)$: the root of $N$;\vspace{-0.5em}
\item ${\cal L}(N)$:  the set of all leaves in $N$;\vspace{-0.5em}
\item ${\cal R}(N)$:  the set of all reticulation nodes in $N$; \vspace{-0.5em}
\item 
${\cal T}(N)$: the set of the root and other tree nodes of outdegree greater than one in $N$; \vspace{-0.5em}
\item ${\cal V}(N)$:  the set of all nodes in $N$ (i.e., ${\cal V}(N)={\cal R}(N)\cup {\cal T}(N) \cup {\cal L}(N)$); \vspace{-0.5em}
\item  ${\cal E}(N)$: the set of all edges in $N$; \vspace{-0.5em}
\item $p(u)$:  the set of the parents of $u\in {\cal R}(N)$ 
or the unique parent of $u\in {\cal T}(N)\backslash \left\{\rho(N)\right\}$;\vspace{-0.5em}
\item $c(u)$:  the set of the children of  
$u\in {\cal T}(N)$ 
or the unique child of  $u\in {\cal R}(N)$;\vspace{-0.5em}
\item ${\cal D}_{N}(u)$:  the \emph{subnetwork} vertex induced by $u\in {\cal V}(N)$ and all the descendants of $u$;
\vspace{-0.5em}
\item 
$N-E$: the spanning \emph{subnetwork} of $N$ with the node set ${\cal V}(N)$ and the edge set 
${\cal E}(N)\backslash E$ for a subset $E\subseteq {\cal E}(N)$;
\vspace{-0.5em}
\item 
 $N-V$:  the \emph{subnetwork} of $N$ with
the node set ${\cal V}(N)\backslash V$ and the edge set $\{ (x, y) \in {\cal E}(N) \;|\; x\not\in V, y\not\in V\}$ for 
a subset $V\subseteq {\cal V}(N)$. 
\end{itemize}

Let $T$ be a phylogenetic tree.  For $S\subseteq {\cal V}(T)$ and $w\in {\cal V}(T)$, $w$ is called the lowest common ancestor (LCA) of the nodes in $S$, denoted $\mbox{lca}(S)$  if 
it is an ancestor of every node in $S$  and any other  ``common" ancestor of the nodes in $S$ is above $w$ in $T$. 

\subsection{The visibility property}

Let $N$ be a network and $u, v\in {\cal V}(N)$.  We say that $u$ is {\it visible} (or stable) on $v$  if every path from the root $\rho(N)$ to $v$ \emph{must} contain $u$ \cite{Huson_Recomb}(also see \cite[p.~165]{Huson_book}). 
In computer science, $u$ is called a dominator of $v$ if $u$ is visible on $v$ \cite{Gero_2013,Leng_74}.
%
%
%

A reticulation node is {\it visible} if it is visible on some leaf.  A network is {\it reticulation-visible}   if every  reticulation node is visible.  It is not hard to see that each reticulation node separates the root from 
some leaves in a reticulation-visible network. 

\begin{figure}[!t]
\begin{center}
\includegraphics[width=0.6\textwidth]{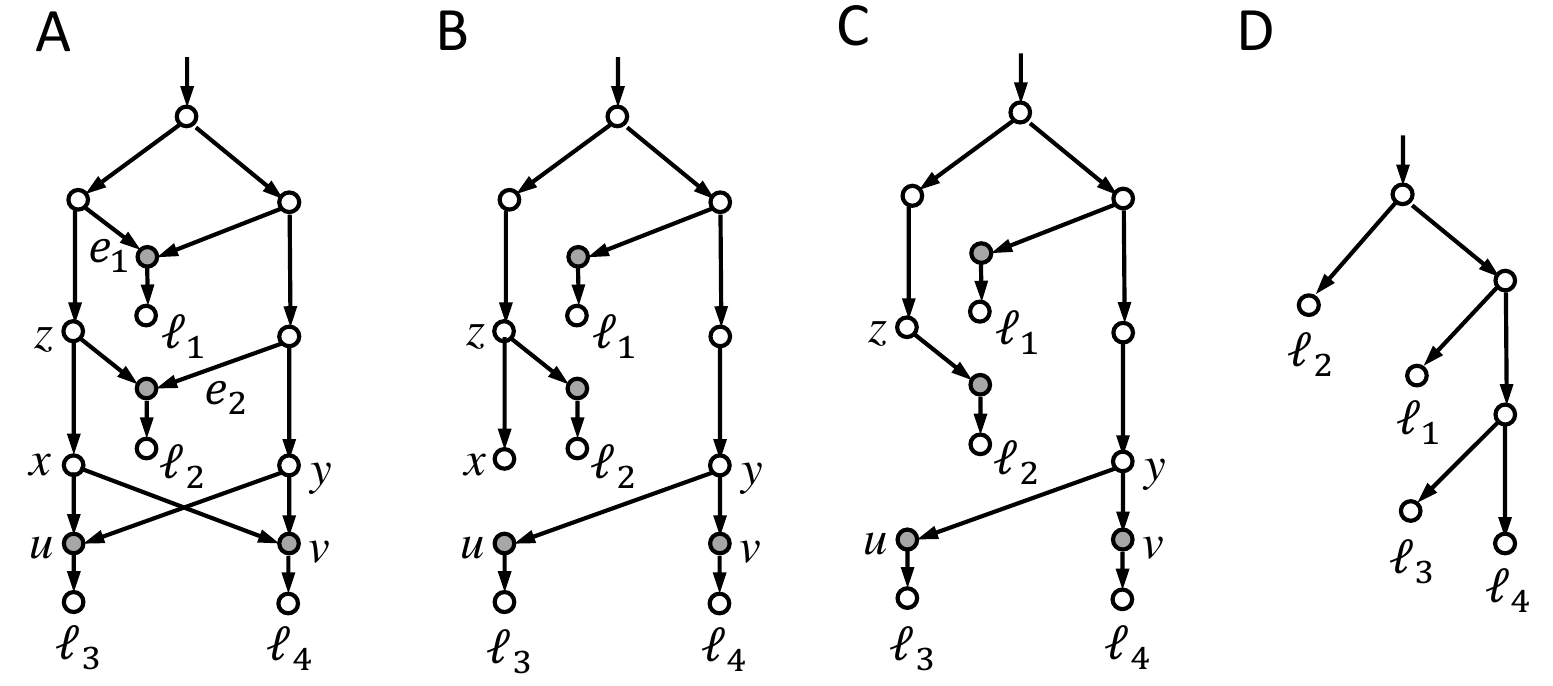}
\end{center}
\vspace*{-0.2in}
\caption{The  network in {\bf A} displays the tree in  {\bf D} through the removal of four edges $e_1, e_2, (x, v)$, and $(x, u)$. The removal of the four edges results in the subtree in {\bf B}, in which $x$ is a dummy leaf; the further removal of  $x$ gives the subtree in {\bf C}, a subdivision of the displayed tree.  Reticulation nodes are represented by  shaded circles. \label{example1}}
\end{figure}

The  network in Figure~\ref{example1}A is reticulation-visible.
Clearly,  all trees are reticulation-visible, as they do not contain any reticulation nodes. In fact, reticulation-visible  networks form a  rather large subclass of networks. The widely studied tree-child networks, galled trees and galled networks are all reticulation-visible \cite{Cardona_09_TCBB, Gusfield_14_Book, Huson_Recomb, Wang_01_JCB}.
Reticulation-visible networks have  two useful properties, as outlined in the following proposition.

\begin{proposition} 
\label{basic_facts}
Let $N$ be a reticulation-visible network and $E\subseteq {\cal E}(N)$.  \vspace{-0.5em}
\begin{itemize}
\item[{\rm (i)}] {\rm ({\bf Reticulation separability})} The child and the parents of a reticulation node are all tree nodes; \vspace{-0.5em}
\item[{\rm (ii)}] {\rm ({\bf Visibility inheritability})}  if  $N-E$ is connected and ${\cal L}(N-E)={\cal L}(N)$,  $N-E$ is also reticulation-visible.
\end{itemize}
\end{proposition} 
\vspace*{-0.5em}

\begin{proofproof}
(i) 
Suppose on the contrary that $u, v\in {\cal R}(N)$ such that $v$ is the child of $u$.  
 Let $w$ be another parent of $v$.   Since $N$ is acyclic, $w$ is not below $v$ and hence is not below $u$.  Since $w$ is not a descendant of $u$,   there is  a path $P(\rho(N), w)$ from $\rho(N)$ to $w$ that does not contain $u$.

We now prove that $u$ is not visible on any leaf by contradiction.
Assume that $u$ is visible on a leaf $\ell$. 
 There is a path $P'$ from $\rho(N)$ to $\ell$ containing $u$. Since $v$ is the only child of $u$,  $v$ appears after $u$ in $P'$. Define $P'[v, \ell]$ to be the subpath of $P'$ from $v$ to $\ell$.  The concatenation of 
 $P(\rho(N),  w)$,  $(w, v)$, and $P'[v, \ell]$ gives a path from $\rho(N)$ to $\ell$. However, this path does not contain
$u$, which is a contradiction. 

\smallskip 
(ii)  
Let $r\in {\cal R}(N-E)$. We assume it is visible on a leaf $\ell$ in $N$. Since ${\cal L}(N-E)={\cal L}(N)$ and $N-E$ is connected, there is at least a path from $\rho(N)$ to $\ell$ in $N-E$.
Any path from  $\rho(N)$ to $\ell$ in $N-E$ is also a path in $N$ and hence must contain $r$. This implies that $\ell$ is visible on $\ell$ in $N-E$.
\end{proofproof}

\subsection{The TCP and CCP}

In a digraph, the {\em suppression} of a node $x$ of indegree  and outdegree one  is a process in which  $x$ is removed and the two directed edges $(y, x)$ and $(x, z)$ incident to $x$  are merged into a directed  edge $(y, z)$. A tree $T'$  is called a {\it subdivision} of another tree $T$ if $T$ can be obtained from $T'$ by a series of suppressions.

Consider a network $N$.  The removal of all but one of the incoming edges for each reticulation node results in a subtree.  However, new (dummy) leaves may or may not exist in the obtained tree. For example, 
the removal of four edges  $e_1, e_2, (x, v)$, and $(x, u)$ in the network given in Figure~\ref{example1}A results in the tree in Figure~\ref{example1}B,  in which $x$ is a new  leaf alongside the  original leaves $\ell_i$ ($1\leq i\leq 4$). If the obtained tree contains dummy leaves,  we will have to remove them and  some of their ancestors to obtain a subtree with the \emph{same} set of leaves as $N$.

\begin{definition}[Tree Containment]
 Let $N$ be a network and $T$ be a phylogenetic  tree such that ${\cal L}(N)={\cal L}(T)$. 
We say that $N$ {\it displays} {\rm (}or contains{\rm )} $T$  if   $E\subseteq {\cal E}(N)$ and  $V\subseteq {\cal V}(N)$ exist such that 
%
{\rm (i)}  $E$ contains all but one of the incoming edges for each 
$u\in {\cal R}(N)$,  and 
{\rm (ii)} $(N-E)-V$  is a subdivision of $T$.
%
\end{definition}

Because of the existence of dummy leaves,  $V$ can be non-empty to guarantee that $N-E-V$ is a subdivision of $T$.
The {\it TCP}  is to  
determine whether or not a  network displays a phylogenetic  tree.
A  bicombining network with $k$ reticulation nodes can display as many as $2^k$ phylogenetic trees.
Hence, for a reticulation network with as many as 32 reticulation nodes \cite[p. 325]{Gusfield_14_Book}, 
a naive exhaustive search is definitely infeasible.  
Therefore, a polynomial-time algorithm is needed for solving the TCP.
%

The set of all the labeled leaves below a node  is called the \emph{cluster} of  the node  in a phylogenetic tree. 
An internal node in a network may have different clusters in different trees displayed in the network. Given a \emph{subset} of labeled leaves $B \subseteq {\cal L}(N)$, $B$ is a {\it soft cluster} in  $N$ if $B$ is the cluster of a node in some tree displayed in $N$. 
%
The {\it CCP}  is to
determine whether or not a  subset $B$ of ${\cal L}(N)$ is a soft cluster in a network $N$. 

The TCP and CCP were both proven to be NP-complete even for binary networks \cite{Kanj_08_TCS}.

\section{A Decomposition Theorem for Reticulation-Visible Networks}
\label{sec4}

In this section, we first present a decomposition theorem for reticulation-visible networks. 
As we shall see later, it plays a vital role  in the study of algorithmic and combinatorial aspects of reticulation-visible networks.


Consider a reticulation-visible network $N$. By Proposition~\ref{basic_facts}, every  reticulation node is only  incident to  tree nodes.  Additionally, in $N-{\cal R}(N)$, each connected component  $C$ (ignoring edge direction) is actually a subtree of $N$ in which the edges are directed away from its root.
Indeed, if $C$ contains two nodes $u$ and $v$ both with indegree zero,  where the indegree is defined over $N-{\cal R}(N)$, and the path between $u$ and $v$ (ignoring edge direction) must contain a node $x$ with indegree $2$, contradicting the assumption that $x$ is a tree node in $N$. 
Hence, the connected components of $N- {\cal R}(N)$ are called the
{\it tree node components} of $N$.  

\begin{figure}[!t]
\begin{center}
\includegraphics[width=0.7\textwidth]{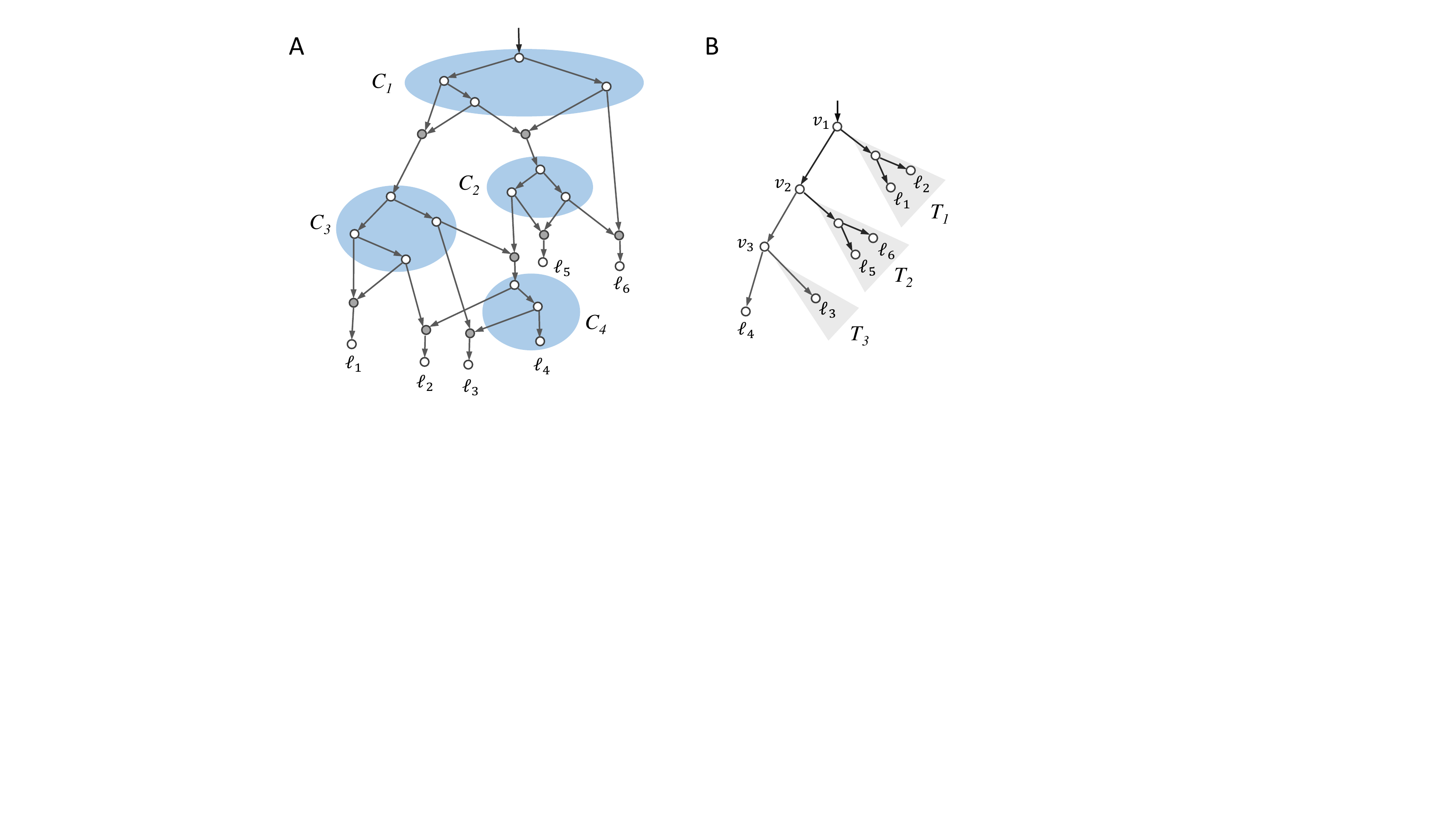}
\end{center}
\caption{({\bf A}) A binary reticulation-visible network with nine tree node components, of which four ($C_1$--$C_4$) are the big ones. 
({\bf B}) A tree considered for containment in the network. When  $C=C_4$ is first selected, we focus on the path from the root $v_1$ to $\ell_4$ in the given tree, where $L_{C} = \{\ell_4\}$,  $s_C = 4$ and $d_C = 3$.\label{example2}
}
\end{figure}

Let $C$ be a tree node  component of $N$ and let ${\cal V}(C)$ denote its node set. $C$
  is called a   {\it single-leaf component}  if ${\cal V}(C)=\{\ell\}$ for some $\ell\in {\cal L}(N)$ and 
  a {\it big} tree node component if $|{\cal V}(C)|\geq 2$.  The  binary reticulation-visible network  in Figure~\ref{example2}A has four big tree node components and five single-leaf components.

 By definition, any two tree node components $C'$ and $C''$ of $N$ are disjoint. We say that $C'$ is \emph{below} $C''$ if  a reticulation node $r$ exists such that   $C'$ is rooted at $r$, whereas a parent of $r$ is in $C''$ in $N$.

\begin{theorem} 
\label{Decomp_Thm}
{\rm ({\bf Decomposition Theorem})}
Let $N$ be a reticulation-visible network with $m$ tree node components 
$C_1, C_2, \dots, C_m$. The following statements are true:
\begin{itemize}
\item[{\rm (i)}] ${\cal T}(N)=\uplus^{m}_{k=1}{\cal V}(C_k)$;
\vspace{-0.5em}
\item[{\rm (ii)}] For each $r\in {\cal R}(N)$,  each of its parents is a tree node in some  $C_i$ and
$c(r)$ is the root of another tree-node component below $C_i$; \vspace{-0.5em}
\item[{\rm (iii)}] For each tree node component  $C_k$, 
$|{\cal V}(C_k)|=1$ if and only if  ${\cal V}(C_k)=\{\ell \}$ for some $\ell\in {\cal L}(N)$ 
{\rm (}i.e., it is  a single-leaf component{\rm )}, 
and if  $|C_k|>1$,  either $C_k$ contains a network leaf or   a reticulation node  exists
 such that  its parents are  all in $C_k$;
 \vspace{-0.5em}
\item[{\rm (iv)}] A big tree node  component $C$ exists, below which there are only  single-leaf  components. 
\end{itemize}
\end{theorem}
\vspace*{-0.5em}
\begin{proofproof}
%
(i) The set equality  follows from the fact  that  the tree node components are different connected components of $N-{\cal R}(N)$ and contain all the tree nodes of $N$.

\smallskip
(ii) Let $r\in {\cal R}(N)$. By the \emph{Reticulation Separability} property (Proposition~\ref{basic_facts}), $c(r)$ and the parents of $r$ are all  tree nodes. Thus by (i), each of them is in a tree node component. 
Furthermore, since 
$c(r)$ is of indegree 0 in $N-{\cal R}(N)$, $c(r)$ must be the root of the tree node component to which it belongs. 


\smallskip
(iii) Let $C$ be a tree node component such that $|C|=1$. Assume  $C=\{u\} \subseteq  {\cal T}(N)\backslash {\cal L}(N)$.  
Since $u$ is the only non-leaf tree node in $C$, $p(u)\in {\cal R}(N)$ and $c(u)\subseteq {\cal R}(N)$. 
Any leaf descendant of $p(u)$ must be below some child of $u$. 
 Let $c(u)=\{c_1, c_2, \dots, c_k\}$.  Since $k$ is finite and $N$ is acyclic, there is a subset $S$
of $i$ children $c_{k_1}, c_{k_2}, \cdots,  c_{k_i}$  such that (i) $c_{k_j}$ is not below any node in $c(u)$ for each $j$, and (ii) each child
in $c(u)$ is either in $S$ or  below some child in $S$. 
For each $j\leq i$,   using the same argument as in the proof of Part (a) of  Proposition~\ref{basic_facts}, we can prove that for each leaf
$\ell$ below $c_{k_j}$, there is path from $\rho(N)$ to $\ell$ that does not contain $p(u)$. 
Since any leaf below $p(u)$ must be below some child in $S$,  $p(u)$ is not visible.
This contradicts the fact that $N$ is reticulation-visible. Therefore,  $|C|=1$ if and only it is a single-leaf component.

\smallskip
Assume that 
$C$ is a big tree node component of $N$ (i.e., $|{\cal V}(C)| \geq 2$). 
Let $\rho(C)$ be the root of $C$.   Since $N$ is reticulation-visible,    the reticulation parent  of $\rho(C)$ and hence $\rho(C)$
itself are both  visible on a network leaf, say $\ell$.  
If $\ell$ is in $C$, the proof is complete. 

 If $\ell$ is not in $C $, we define ${\cal X}=\{ r\in {\cal R}(N) \;|\;  p(r)\cap C\neq \emptyset \;\;  \text{and} \;\; \ell \mbox{ is below $r$}\}$. Clearly, ${\cal X}$ is finite.
For any $r', r''\in {\cal X}$, we write $r'\prec_{\cal X} r''$ if $r'$ is below $r''$; in other words, there is a direct path from $r''$ to $r'$.  Since $\prec_{\cal X}$ is transitive and   $N$ is acyclic, ${\cal X}$ contains a maximal element 
$r_m$  with respect to $\prec_{\cal X}$. Let $p(r_m)=\{p_1, p_2, \cdots, p_k\}$.  Since $r_m\in {\cal X}$, we may assume 
that $p_1\in {\cal V}(C)$. If $p_{k_0} \not\in {\cal V}(C)$ for some $1<k_0\leq k$,  $p_{k_0}$ is not below any node in $C$, as $N$ is acyclic and $r_m$ is maximal under $\prec_{\cal X}$. Hence,  there is a path $P$ from $\rho (N)$ to $p_{k_0}$ that does not contain any node in 
$C$.   Since $\ell$ is a descendant of $r_m$,  $P$ can be extended into a path from $\rho (N)$ to $\ell$ that does not contain $\rho(C)$. This contradicts the statement that $\rho(C)$ is visible on $\ell$. Therefore, the parents of $r_m$ are all in $C$.

%

\smallskip
(iv)
This is derived from the fact that $N$ is acyclic and finite. 
\end{proofproof}

Let $N$ be a binary reticulation-visible  network.
Since $N$ is a directed acyclic graph and has, at most, $8\,|{\cal L}(N)|$ nodes \cite{Gunawan_2015},  we can determine the tree node components using the breadth-first search technique in $O(|{\cal L}(N)|)$ time.
Additionally, 
a topological ordering of its nodes can also  be found in $O(|{\cal L}(N)|)$ time. 
Using this topological ordering, we can derive another topological ordering for the big tree node components, with which we can identify the lowest tree node component described in Theorem~\ref{Decomp_Thm}(iv) in constant time.

 For non-binary networks,  the above processes for finding all the big tree node components and determining the lowest ones take 
$O(|{\cal V}(N)|+|{\cal E}(N)|)$ time.

\section{Galled Networks}

We shall apply  the decomposition theorem to bound the number of nodes in a galled network in this section.
A { bicombining} network is {\it galled} if each reticulation node $r$ has an ancestor  $u$ such that  two internal node-disjoint paths exist  from $u$ to $r$ in which all nodes except $r$ are tree nodes. 

Let N be a reticulation-visible network.  A $r\in {\cal R}(N)$ is {\it inner}  if its parents are all in the same tree node component of $N$; it is called a {\it cross-reticulation} otherwise.

\begin{theorem}
Let $N$ be a bicombining network. 
 $N$  is galled if and only if it is reticulation-visible and every reticulation is inner.
\end{theorem}
\begin{proofproof} 
(Necessity)  Let $N$ be  galled.  $N$ is reticulation-visible \cite[p. 165]{Huson_book}. Assume, on the other hand, that  $N$ contains  a cross-reticulation $r$. By definition, 
the parents of $r$ are in different tree node components. 
Assume that $p_1$ and $p_2$ are  two parents of $r$ in different tree node components.  Since $N$ is acyclic, we may assume that {\it  $p_1$ is not below 
$p_2$}.  
Let $C_{p_i}$ be the tree node components containing $p_i$ for $i=1, 2$. 
We now consider the parent $r'$ of  $\rho(C_{p_2})$.  First, $r'$ is a reticulation node. Second, $p_2$ is below $r'$ and hence $r$ is also below $r'$. 
However, we can reach $r$ from $p_1$ using a single edge without passing through $r'$, contradicting 
the Separation Lemma for galled networks \cite[p. 163]{Huson_book}.   

(Sufficiency)  Let $N$ be a bicombining reticulation-visible network such that each reticulation is inner.  For each  reticulation node $r$, by definition, its parents $p'_r$ and $p''_r$ are both in the same tree node component $C$. Since $C$ is a subtree of $N$, $\mbox{lca}(p'_r, p''_r)$ is also a tree node in $C$ and hence  two 
internal node-disjoint paths from $\mbox{lca}(p'_r, p''_r)$ to $r$ exist in which all but $r$ are tree nodes.  Therefore, $N$ is galled.
\end{proofproof}

\begin{corollary}
\label{bound} If $N$ is a galled network with $n$ leaves, then

 {\rm (i)}   $| {\cal R} (N)| \leq 2(n-1)$;

 {\rm (ii)}   $|{\cal V}(N)| \leq 6n-5$.
\end{corollary}
\vspace*{-0.5em}
\begin{proofproof}
(i) Let $N$ be a galled network with $n$ leaves. 
We will consider the decomposition of $N$ into tree node components. 
Since the root of each tree node component is either $\rho(N)$ or the unique child of a reticulation, the following holds:
\begin{eqnarray}
\label{key_eqn}
|{\cal R}(N)| =\mbox{(no. of tree node components in $N$)} -1.
\end{eqnarray}

We have proved that $N$ contains only inner reticulation nodes, the parents of which are all in the same tree node component.  Therefore,  all the tree node components are connected in a tree structure. Specifically, if $G$ is 
the graph for which the nodes are the tree node components and in which 
a node $X$ is connected to another $Y$ by an edge if the tree node components corresponding to them are separated by 
a reticulation node between them, then $G$ is a rooted tree.

Consider a leaf $l$  in $G$.  Since there is no reticulation node below  tree-node component represented by $l$, 
it must contain at least one leaf.

In addition,  $G$ may contain internal nodes of degree two. Assume that $C_x$ is a tree node component represented by
an internal node $x$ of degree two in $G$.  Here, there is only an inner reticulation $r$ below $C_x$. Since there are no multiple edges between any pair of nodes, $C$ must contain a network leaf. 

Let  $n'$ be the number of network leaves contained in the tree node components corresponding to nodes of degree two in $G$.
 $G$ has $n-n'$ leaves at most and hence $n-n'-1$ internal nodes of degree greater than two. 
 Therefore,  the number of nodes in  $G$ is, at most,   $n'+(n-n')+(n-n'-1)\leq 2n-1$. In other words, $N$ contains $2n-1$ tree node components at most.  By Eqn.~(\ref{key_eqn}),
			$|{\cal R}(N)| \leq 2(n-1).$

(ii)    For a node $u$ in $N$, we use $d_{i}(u)$ and $d_{o}(u)$ to denote 
     the indegree and outdegree  of $u$, respectively.  
     Since $N$ is a bicombining network, each reticulation node has indegree two and outdegree one.    Hence:  
\begin{eqnarray}
    n+|{\cal T}(N)\backslash \{\rho(N)\}| + 2|{\cal R}(N)| =\sum_{v\in {\cal V}(N)}d_i(v)= \sum_{u\in {\cal V}(N)}d_o(u) = \sum_{u\in {\cal T}(N)} d_{o}(u) + |{\cal R}(N)|,
\end{eqnarray} 
where ${\cal T}(N)$ is the set of the nodes of outdegree greater than one, including the root. This 
 implies that:
   $$|{\cal T}(N)| \leq  \sum_{u\in {\cal T}(N)} (d_{o}(u)-1) = |{\cal R}(N)|  + n-1\leq 3(n-1).$$
 Therefore, $|{\cal V}(N)|= n + |{\cal T}(N)| + |{\cal R}(N)| \leq 6n -5.$
\end{proofproof}

\section{A Quadratic-time Algorithm for the TCP}
\label{sec5}

In this section, we shall  present a dynamic programming algorithm for the TCP that takes quadratic-time. 

\subsection{The rationale behind our algorithm}

The TCP has been known to be solvable in polynomial time only  for tree-child networks \cite{van_Iersel_2010_IPL} and the so-called nearly-stable networks \cite{Gunawan_2015}. 
In a tree-child network,  each reticulation node is essentially connected to a leaf by a path consisting of only tree nodes.   In a nearly-stable network,  each  child  of a node is visible if the node is not visible. 
Because of the simple local structure around a reticulation node in such a network, one can determine whether or not the network displays a phylogenetic tree by examining the reticulation nodes one by one.
However, any approach that works on reticulation nodes one by one  is not powerful  enough for solving the TCP for a reticulation-visible network with the structure shown in Figure~\ref{example1}, which could have many reticulation nodes above the parents of two reticulation nodes at the bottom.  We have to deal with the whole set of reticulation nodes simultaneously for a reticulation-visible network of this kind.

 Our algorithms for the TCP and CCP  rely
primarily on the decomposition theorem (Theorem~\ref{Decomp_Thm}). 
Recall that this theorem says  that in a reticulation-visible network,
all non-reticulation nodes can be partitioned into a collection of disjoint connected tree node components, each
having at least {\it  two nodes} if it contains a non-leaf tree node (Figure~\ref{example2}).  Most importantly,  each component  {\it contains}  either a network leaf or  all the parents of a reticulation node.

The  topological property uncovered by this theorem allows us to solve the TCP by  the divide-and-conquer
approach:  We work on the tree node  components one by one in a bottom-up fashion.  When working on a tree node
component, we construct a tree in which multiple leaves may have the same label and 
 compare it with the input tree to decipher all the reticulation nodes immediately below it. 

\subsection{The algorithm}

%

\begin{figure}[!t]
\begin{center}
\includegraphics[width=0.6\textwidth]{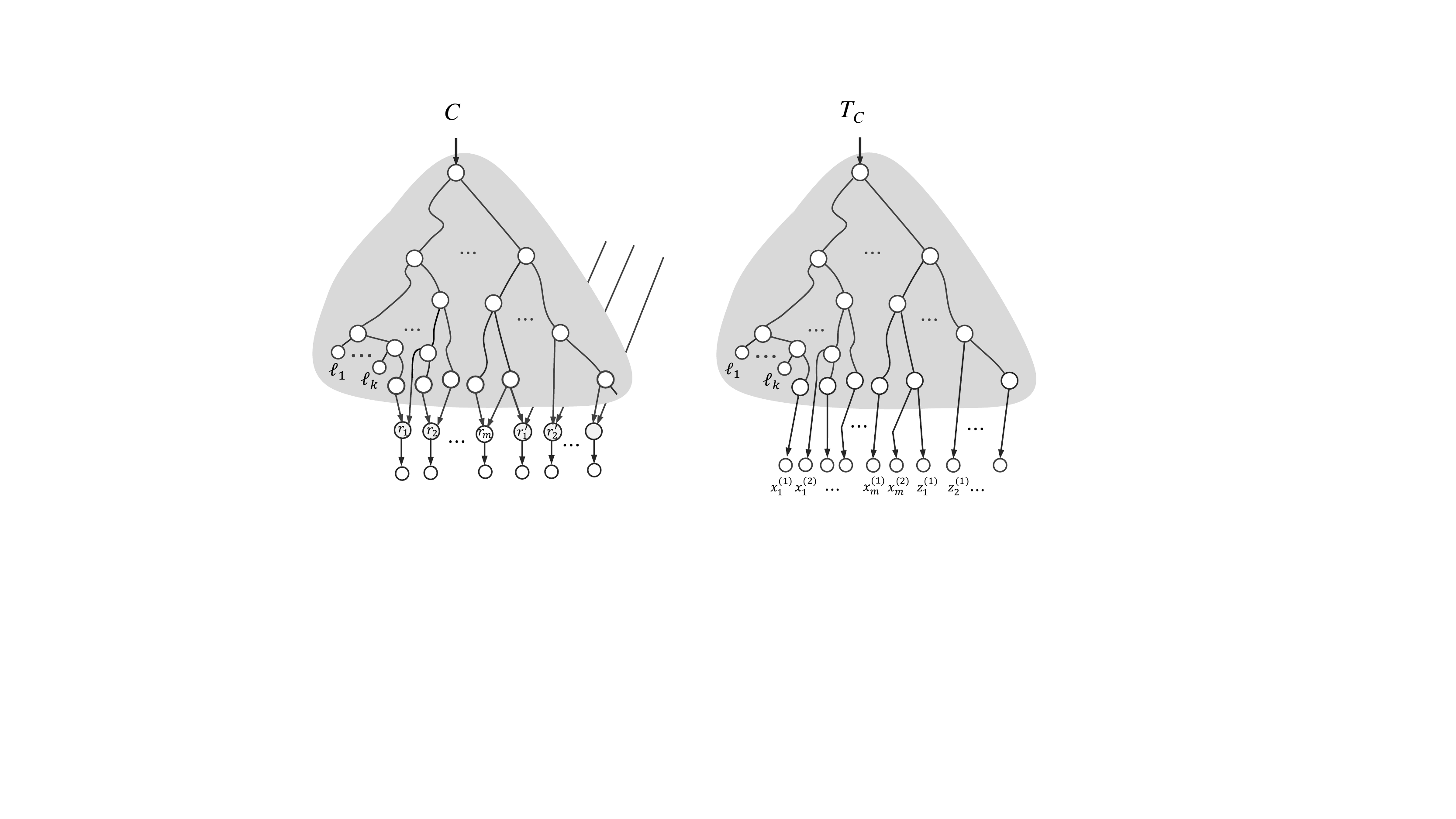}
\end{center}
\caption{Illustration of the lowest tree node component $C$  in a reticulation-visible network and the corresponding tree $T_C$ 
constructed for computing $d_C$ in Proposition~\ref{Prop_55}. Here, each reticulation node is of indegree two. \label{Fig3}
}
\end{figure}

By Theorem~\ref{Decomp_Thm}(iv), a ``lowest'' big tree-node component $C$ exists,
below which there are only (if any) single-leaf components (Figure~\ref{Fig3}). 
We assume that $C$ contains
$k$ network leaves
\begin{eqnarray}
  \ell_1, \ell_2, \cdots, \ell_k,
\end{eqnarray}
 and that below $C$, there are $m$ inner reticulations: 
\begin{eqnarray} \mbox{IR}(C)=\{r_1, r_2, \cdots, r_m\},
\end{eqnarray} and $m'$ cross-reticulations:
\begin{eqnarray}
\mbox{CR}(C)=\{r'_1, r'_2, \cdots, r'_{m'}\}.
\end{eqnarray}
Since  $C$ is a big tree node component, by Theorem~\ref{Decomp_Thm}(iii),  it has two or more nodes, implying that $k+m+m'\geq 2$.   
%
%

Let $\rho(C)$ denote the root of $C$. We further define:
\begin{equation}
\label{eq0}
L_C = \left\{\,\ell_1, \ell_2, \dots, \ell_k, c(r_1), c(r_2), \cdots, c(r_m)\right\}.\end{equation}
 By Theorem~\ref{Decomp_Thm}(iii), 
$k+m\geq 1$ (i.e., $L_C$ contains at least one leaf of $N$).   

\begin{proposition}
\label{Prop50}
The root $\rho(C)$ of $C$ is visible on each leaf  $\ell \in L_C$ in $N$.  
\end{proposition}
\begin{proofproof}
For each $\ell_i \in L_C$ ($1\leq i\leq k$), each path from $\rho(N)$ to $\ell_i$ must pass $\rho(C)$ before reaching $\ell_i$, as there are only tree nodes in the unique path
from $\rho(C)$ to $\ell_i$. 

For each $c(r_i)\in L_C$ ($1\leq i\leq m$), any  path $P$ from $\rho(N)$ to $c(r_i)$ must contain $r_i$. Since the parents of $r_i$ are all in $C$, $P$ must contain $\rho(C)$. 

Taken together,  the facts imply that $\rho(C)$ is visible on each network leaf in $L_C$.  
\end{proofproof}

We select $\ell \in L_C$.
Since $T$ has the same leaves as $N$, $\ell \in {\cal L}(T)$ and 
a unique path $P_T$ exists  from  $\rho(T)$ to $\ell$ in $T$.  Let:
\begin{equation}
\label{eq1}
P_T:\; v_1, v_2,  \dots, v_{t}, v_{t+1},
\end{equation} 
where $v_1=\rho(T)$ and $v_{t+1}=\ell$.
  $T-P_T$ is then the union of $t$ disjoint subtrees $ T_1, T_2,  \dots, T_{t}$, where $T_i$ ($1\leq i\leq t$)  is the subtree branching off from $P_T$ at $v_i$  (Figure~\ref{example2}B). For the sake of convenience, we also consider the single leaf $\ell$ as a subtree, written as $T_{t+1}$. 
Define:
 \begin{equation}
 \label{eq2}
 s_C=\min \{s  \;|\; {\cal L}(T_s)\cap L_C \neq \phi \}.
\end{equation} 
 Since $\ell \in {\cal L}(T_{t+1})\cap L_C$,  $s_C$ is well defined. For example,  $s_C=4$ for the instance given in Figure~\ref{example2}.

\begin{proposition}
\label{Prop_51}
The index $s_C$ can be computed in  $O(|{\cal L}(N)|)$ time.
\end{proposition}
\vspace*{-0.5em}

\begin{proofproof}  
 Since $T$ is a binary tree with the same set of labeled leaves as the network $N$,
$T$ has $2|{\cal L}(N)|-1$ nodes and $2|{\cal L}(N)|-2$ edges.  For each $x\in {\cal V}(T)$, we define a flag variable $f_x$ to indicate whether the subtree below $x$ contains a network leaf in $L_C$ or not.   We first traverse $T$ in the post-order: \vspace*{-0.5em}
\begin{itemize}
\item For a leaf $x\in {\cal L}(T)$, $f_x=1$ if $u\in L_C$ and 0 otherwise. \vspace*{-0.5em}
\item For a non-leaf node $x$ with children $y$ and $z$, $f_x=\max\{f_y, f_z\}$. \vspace*{-0.5em}
\end{itemize} 
By definition, 
$s_C=\min \{ i\;|\; f_{\rho(T_i)}=1\}.$
Obviously, the whole process computes $s_C$ in $O(|{\cal L}(T)|)=O(|{\cal L}(N)|)$ time.
\end{proofproof}

\begin{proposition}
\label{Prop52}
If $N$ displays $T$, 
then ${\cal D}_{T}\left(v_{s_C}\right)$ is displayed in ${\cal D}_N\left(\rho(C)\right)$. 
\end{proposition} 
\vspace*{-0.5em}
\begin{proofproof} 
When $s_C = t+1$,  ${\cal D}_{T}\left(v_{s_C}\right)=\{\ell \}$ and thus the statement is true. 

When $s_C < t+1$, by the definition of $s_C$,   a network leaf $\ell'$  exists in $ {\cal L} (T_{s_C})\cap L_C$ such that $\ell'\neq \ell$. 
If $N$ displays $T$,   a subdivision $T'$ of $T$ exists in $N$.  By Proposition~\ref{Prop50},   $\rho(C)$ is visible on both $\ell$ and $\ell'$. 
This implies that   $\ell$ and  $\ell'$ are both below   $\rho(C)$ in $T'$. Since $T'$ is a tree,  $\mbox{lca}_{T'}(\ell, \ell')$, the LCA of $\ell$ and $\ell'$,  is also below $\rho(C)$ in $T'$.  Since  $\mbox{lca}_{T'}(\ell, \ell')$ is the node in $T'$ that corresponds to $v_{s_C}$,  
 the subnetwork of $T'$ below $\mbox{lca}_{T'}(\ell, \ell')$ is a subdivision of  ${\cal D}_{T}\left(v_{s_C}\right)$ (i.e., 
 ${\cal D}_N\left(\rho(C)\right)$ displays ${\cal D}_{T}\left(v_{s_C}\right)$).
\end{proofproof}

If $N$ displays $T$,  then $C$ may display a subnetwork of $T$ that properly contains ${\cal D}_{T}\left(v_{s_C}\right)$. In other words, it may display a subtree  ${\cal D}_{T}\left(v_j\right)$ for some $j < s_C$.
We define:
\begin{equation}
\label{eq3}
 d_C=\min \,\left\{ j \; |\; {\cal D}_{T}\left(v_j\right) \mbox{ is displayed in ${\cal D}_N \left(\rho(C)\right)$} \,\right\}
\end{equation}
For example, $d_C=3$ for the instance  given in Figure~\ref{example2}.

\begin{proposition}
\label{Prop53}
If $N$ displays $T$,  $T$ has a subdivision $T''$  in $N$  such that  
the node corresponding to $v_{d_C}$ is in $C$. 
\end{proposition}
\vspace*{-0.5em}
\begin{proofproof}
Let $N$ display $T$.  Assume that $T'$ is a subdivision of $T$ in $N$ in which   the node  corresponding to $v_{d_C}$ is $u$. 
If $u$ is  in $C$,  the proof is complete. 

Assume that $u$ is not in $C$.  
 Since $\ell$ is below $v_{d_C}$ in $T$, $\ell$ is below $u$ in $T'$.  
By Proposition~\ref{Prop50},  $\rho(C)$ is visible on  $\ell$, and $\rho(C)$ is contained in the unique path $P$ from the network root to $\ell$ in $T'$. 
  Since $u$ is in $P$ but not in $C$,  $\rho(C)$ is  below $u$ in $P$.  Let $P'$ be the subpath of $P$ from $u$ to 
$\rho(C)$.

By assumption, ${\cal D}_{T}(v_{d_C})$ is displayed in ${\cal D}_N(\rho(C))$. It has a subdivision $T^*$ in ${\cal D}_N(\rho(C))$. Let $u'$ be the node in $T^*$ corresponding to 
$v_{d_C}$. 
Let  $P''$  be the path  from  $\rho(C)$ to $u'$ in $C$. 
Since the subtree below $u'$ in $T^*$ and the subtree below $u$ in $T'$ are the subdivisions of the subtree below $v_{d_C}$ in $T$,  $T'-{\cal D}_{T'}(u) + P' +P''+ 
 {\cal D}_{T^{*}}(u')$ is also 
 a subdivision of $T$ in $N$, in which $u'$ is the node in $C$ corresponding  to $v_{d_C}$. Here, $G+H$ is the graph with the same node set as $G$ and the edge set is the union of $E(G)$ and $E(H)$ for graphs $G$ and $H$ such that $V(H) \subseteq V(G)$.
\end{proofproof}

To compute $d_C$ as defined in Eqn.(\ref{eq3}) in linear-time,
we create a tree $T_C$ from $C$ by attaching a new leaf, which has the same label as $c(r)$,  below every parent node in $p(r)\cap {\cal V}(C)$  for each $r\in \mbox{IR}(C)\cup \mbox{CR}(C)$. 
 In other words, 
 $T_C$ has  the node set:
\begin{eqnarray}
\label{eq44}
 {\cal V}(T_C) &=&{\cal V}(C)  \nonumber \\
   & &  \cup \; \{ x^{(i)}_r \;|\;
   r\in \mbox{IR}(C)  \;\& \; 1\leq i\leq |p(r)| \} \nonumber \\
   && \cup \; \{ z^{(i)}_r \;|\; r\in \mbox{CR}(C)  \;\& \; 1\leq i\leq k \}, 
\end{eqnarray}
where $k=|p(r)\cap {\cal V}(C)|$ and $x^{(i)}_r$  and $z^{(i)}_r$ are new leaves with the same label as $c(r)$ for each $r\in \mbox{IR}(C)\cup \mbox{CR}(C)$, and with the edge set
\begin{eqnarray}
\label{eq55}
  {\cal E}(T_C)&=&{\cal E}(C)  
    \cup \; \{ 
 (u^{(i)}_r, x^{(i)}_r) \;|\;
  r\in \mbox{IR}(C) \mbox{ s.t. }  p(r)=\{u^{(i)}_r \;|\; 1\leq i\leq  |p(r)|\}\} \nonumber \\ 
& &\cup \; \{ (v^{(i)}_r, z^{(i)}_r) \;|\;r\in \mbox{CR}(C)
\mbox{ s.t. } p(r)\cap {\cal V}(C)=\{v^{(i)}_r \;|\; 1\leq i\leq  k  \}\}.
\end{eqnarray}
$T_C$ has $1+\sum_{v\in {\cal V}(C)}d_o(v)$ nodes at most, where $d_o(v)$ is the outdegree of $v$ in $N$. 
For bicombining reticulations,  $T_C$ is illustrated in Figure~\ref{Fig3}.

For each $r\in \mbox{IR}(C)$, $T_C$ contains multiple leaves with the same label as $c(r)$. 
To detect whether or not $\cD_T\left(v_j\right)$ is displayed in $\cD_N\left(\rho(C)\right)$, 
we have to consider which of these leaves will be removed. 
Such leaves will be referred to as the {\it ambiguous} leaves. 
%
We define:
\begin{eqnarray}
\label{ar-def}
A_r = \{x^{(i)}_r  \;|\; 1\leq i\leq |p(r)| \}
\end{eqnarray}
for $r\in \mbox{IR}(C)$ and
\begin{eqnarray}
\label{ar-def}
A(T_C)=\cup _{r\in \mbox{\scriptsize IR}(C)} A_{r}.
\end{eqnarray} 
$A(T_C)$ is the set of ambiguous  leaves in $T_C$.

For each $r\in \mbox{CR}(C)$, $T_C$ may also have multiple leaves with the same label as $c(r)$. 
Unlike ambiguous leaves,  all these leaves can be removed to obtain a subtree of $N$.  Such leaves are called 
{\it optional} leaves. We use $O(T_C)$ to denote the set of optional leaves in $T_C$.

Determining whether or not a subtree of $T$ is displayed in $C$ is very similar to the problem studied in \cite{Cui_Zhang}. 
Here, we shall  develop a dynamic programming algorithm for the task.  More precisely, we shall compute the following set $S_u$ of nodes in $T$:
$$ S_u =\{ x \in {\cal V}(T) \;|\;  {\cal D}_{N}(u) \mbox{ displays  the subtree ${\cal D}_{T}(x)$ where  $u$ corresponds to
$x$}\}$$
for each node $u$ in $T_C$.
Here,  $x\in S_u$ if and only if each leaf  $\ell$ below $x$ in $T$ is uniquely mapped to a leaf $\ell'$ below $u$ in $T_C$ such that,
for $I=\{\ell' \in {\cal L}({\cal D}_{T_C}(u)) \;|\; \ell \in {\cal L}({\cal D}_{T}(x))\}$, the following are true: \vspace*{-0.5em}
\begin{itemize}
\item[(i)] the  subtree of  $T_C$ over $I$ is a subdivision of ${\cal D}_{T}(x)$, \vspace*{-0.5em}
\item[(ii)] $I\cap \{\ell_1, \ell_2, \cdots, \ell_k\} =  {\cal L}({\cal D}_{T_C}(u)) \cap \{\ell_1, \ell_2, \cdots, \ell_k\}$, and \vspace*{-0.5em}
\item[(iii)] $|{\cal L}({\cal D}_{T_C}(u)) \cap A_r| < |A_r|$ for each $r\in \mbox{IR}(C)$ such that $c(r)\not\in {\cal L}({\cal D}_{T}(x))$. \vspace*{-0.5em}
\end{itemize} 
The reason for (ii) is that we cannot eliminate a network leaf in $C$ when we remove the edges entering a reticulation node to form a subtree of $N$.
The reason for (iii) is that we must  keep  exactly one incoming edge of each $r\in \mbox{IR}(C)$ to form a subtree of $N$.

We introduce a Boolean variable 
$f_{ux}$ to indicate whether or not $x\in S_u$  and  a set variable $M_{ux}$. That is, 
\begin{eqnarray}
f_{ux}=\left\{ \begin{array}{ll}
        1 & \mbox{if ${\cal D}_{T_C}(u)$  displays   ${\cal D}_{T}(x)$,}\\
        0 & \mbox{otherwise.}
   \end{array}
 \right. \label{fux_def}
\end{eqnarray}
%
%

We recursively compute
$f_{ux}$ for each $u$ and $x$  by traversing both $T_C$ and $T$
in the post-order.  To compute $f_{ux}$ in linear time for a fixed $x$, we have to identify which $u$ does not satisfy Conditions (ii) and (iii) mentioned in the last paragraph.  Let $x$ be an internal node in $ T$ such that there are at least two leaves below it (i.e., $|{\cal L}({\cal D}_{T}(x))|>1$).  

\begin{proposition}
 \label{prop564}
All the nodes in $V_{\mbox{\rm lca}}=\{\mbox{\rm lca}_{T_C}(A_r) \;|\;  r\in \mbox{IR}(C)\}$  can be determined in
 $O(|{\cal E}(T_C)|)$ time. 
\end{proposition}
\begin{proofproof} 
 We first pre-process $T_C$ in $O(|{\cal E}(T_C)|)$ time so that for any two nodes $u$ and $ v$ in $T_C$, $\mbox{lca}_{T_C}(u, v)$ can be found in $O(1)$ time \cite{Farach, Tarjan}.

Initially, each LCA node is undefined. We visit  all leaves in $T_C$ in the post-order.  When visiting a leaf $\ell$ that is an ambiguous leaf in $A_r$ for 
$r\in \mbox{IR}(C)$,  
  we set $\mbox{lca}_{T_C}(r)=\ell$  if  
$\mbox{lca}_{T_C}(r)$ is undefined,  and $\mbox{lca}_{T_C}(r)=\mbox{lca}_{T_C}(\mbox{lca}_{T_C}(r), \ell)$ otherwise.  Since each LCA operation takes $O(1)$ time, 
the whole process takes $O(|{\cal E}(T_C)|)$ time. 
\end{proofproof}
 
\begin{proposition}
\label{prop565}
 {\rm  (i)} Let $\ell$ be a leaf in $T_C$ that is  neither ambiguous nor optional.  If $\ell \notin {\cal L}({\cal D}_{T}(x))$,  $f_{ux}=0$  for any   $u$  in the unique  path from $\rho(C)$ to $\ell$ in $N$. 

 {\rm (ii)}  For each $r\in \mbox{IR}(C)$ such that $c(r) \notin {\cal L}({\cal D}_{T}(x))$,   
$f_{ux}=0$ for  any $u$  in the path from $\rho(C)$ to $\mbox{\rm lca}_{T_C}(A_r)$ inclusively in $N$. 
\end{proposition}
\begin{proofproof}
 (i) Since $\ell$ is neither ambiguous nor optional, all the nodes in the path from $\rho(C)$ to $\ell$ appear in any subtree $T$ of $N$. 
   Since $\ell\not\in {\cal L}({\cal D}_{T}(x))$,  $f_{ux}=0$ for any $u$ in  the unique path from $\rho(C)$ to $\ell$ in $N$. 

 (ii)  Let $u$ be a node in the path from $\rho(C)$ to $\mbox{\rm lca}_{T_C}(A_r)$ inclusively in $N$. For each $r\in \mbox{IR}(C)$, $A_r$ contains at least two ambiguous leaves and thus $\mbox{lca}_{T_C}(r)$ is an internal  node in $C$. Any subtree $T$ of $N$ contains exactly  one incoming edge of $r$  below $\mbox{lca}_{T_C}(r)$.  This implies that  $c(r)$ must appear in any subtree displayed below $u$.   Since $c(r) \notin {\cal L}({\cal D}_{T}(x))$, ${\cal D}_{T}(x)$ cannot be displayed at $u$. 
\end{proofproof} 

Let $\bar{T}_x$ be the spanning subtree of $T_C$ over $\{ \ell \in {\cal L}(T_C) \;|\; 
\ell \not\in A(T_C)\cup O(T_C)\cup {\cal D}_{T}(x)\} \cup V_{\mbox{lca}} \cup \{\rho(C)\}$.
$\bar{T}_x$ is a subtree on the top of $T_C$ with the same root $\rho(C)$. 

By Propositions~\ref{prop564} and \ref{prop565}, a node in $T_C$ satisfies Conditions (ii) and (iii) if and only if it is not in $\bar{T}_x$.  Therefore, we have the following fact.

\begin{proposition}
\label{prop566}
For $u \in {\cal V}(T_C)$,   $f_{ux}=1$ if and only if  $u\not\in {\cal V}(\bar{T}_x)$ and one of the following two conditions is true:
\vspace{-0.5em}
\begin{itemize}
\item[{\rm (a)}] a child $v\in c(u)$ exists such that $f_{vx}=1$; \vspace{-0.5em}
\item[{\rm (b)}] $x$ has the children $y$ and $z$ and distinct $v'\in c(u)$ and $v''\in c(u)$ exist such that 
  $f_{v'y}=1$ and $f_{v''z}=1$. \vspace{-0.5em}
\end{itemize}
\end{proposition}
%

\begin{proposition}
\label{Prop_55}
There is an algorithm that takes $T_C$ and $T$ as input and computes $d_C$ in Eqn.~(\ref{eq3}) in $O(|{\cal E}(T_C)|\cdot|{\cal V}(T)|)$ time. 
\end{proposition} 
\vspace*{-0.5em}
\begin{proofproof}
Our dynamic programing algorithm traverses $T$ in the post-order: for each $x$,  compute $f_{ux}$ for all $u$ in $T_C$ in four steps:

 Step 1.  Pre-process $T_C$ so that the LCA of any two nodes can be found in $O(1)$ time in $T_C$.

 Step 2.  Traverse the leaves in $T_C$ to compute the nodes in $V_{\mbox{lca}}$. 

 Step 3.   Mark the nodes in the subtree  $\bar{T}_x$ of $T_C$. For each  leaf $\ell \not\in A(C) \cup O(C)\cup {\cal L}({\cal D}_{T}(x))$, mark the nodes in the path from $\rho(T_C)$ to $\ell$.   For each $r\in \mbox{IR}(C)$ such that $c(r) \not\in {\cal L}({\cal D}_{T}(x))$, mark the nodes in the path from $\rho(T_C)$ to $\mbox{lca}_{T_C}(A_r)$ inclusively. 

Step 4.   Traverse $T_C$ in a bottom-up manner. For each node $u$,  if it is {\it unmarked} in Step 3, compute $f_{ux}$ using the formulae given in Table~\ref{table2}; if  it is marked in Step 3, $f_{ux}=0$.

The correctness of the above computation process  follows from Propositions~\ref{prop565} and \ref{prop566}. 
Step 1 takes constant time \cite{Farach, Tarjan}.
By Proposition~\ref{prop564}, Step 2 can be done in $O(|{\cal E}(T_C)|)$ time. 

Two paths from $\rho(T_C)$ to nodes in $V_{\mbox{lca}}$   may have a common part starting at $\rho(T_C)$. 
We mark the nodes in each of these paths in a bottom-up manner: whenever we reach a marked node, we stop the marking process in the current path.
In this way, each marked node is visited twice at most and hence Step 3 can be executed in $O(|{\cal E}(T_C)|)$ time. 

In Step 4, at a node $u$, we simply need to check whether or not $f_{vx}=1$ for each child $v$ of $u$ and
whether or not $f_{vy}=1$ for each child $y$ of $x$ and each child $v$ of $u$. Since $x$ has only two children, 
Step 4 takes $\sum_{u\in {\cal V}(T_C)} O(|c(u)|)=O(|{\cal E}(T_C)|)$ time.  

Taken together, these facts imply that  our algorithm takes $\sum_{x\in {\cal V}(T)}O(|{\cal E}(T_C)|)=O(|{\cal N}(T_C)|\cdot|{\cal V}(T)|)$.

After we know the values of $f_{ux}$ for every $u$ in $T_C$ and every $x$ in $T_c$, 
we can compute $d_C$ such that ${\cal D}_{T}(v_j)$ is displayed in 
${\cal D}_{N}(\rho(C))$ as:
$$ d_C= \min _{1\leq j\leq t+1} \{ j \;|\; f_{u v_j}=1 \mbox{ for some } u\in {\cal  V}(T_C) \}.$$
\end{proofproof}

\begin{table}[!t]
\begin{center}
{\small 
\begin{tabular}{c|l}
\toprule
  $x\in {\cal L}(T)$? & $f_{ux}$, where $u$ is an unmarked node in $T_C$\\
\midrule
  Yes  & Case 1:   $f_{vx}=0$ for each $v\in c(u)$\\
   &~~$f_{ux}=0$;\\
  & \\
& Case 2:  $f_{vx}=1$ for some $v\in c(u)$\\
& ~~ $f_{ux}=1$;\\
\midrule
No.   &  Case 3:   $f_{vx}=0$ for each $v\in c(u)$, and \\
 $c(x)=\{y, z\}$ & ~~~~~~~~~ no distinct  $v'$ and $v''$ in $c(u)$ exist such that
  $f_{v'y}=1$ and $f_{v''z}=1$;\\
&~~$f_{ux}=0$;\\
& \\
    &   Case 4:  $f_{vx}=1$ for some $v\in c(u)$, or \\
   &~~~~~~~~~ $f_{v'y}=1$ and $f_{v''z}=1$ for some  $v'\neq v''$ in $c(u)$\\
& ~~$f_{ux}=1$;\\
\bottomrule
\end{tabular} 
} 
\end{center}
\caption{The update rules for computing  $f_{ux}$ for an internal  node $u$ in $T_C$. \label{table2}}
\end{table}

\newpage 
Based on the facts presented above, we obtain the following  algorithm for the TCP. \vspace{1em} 

\begin{center}
\begin{tabular}{l}
\toprule
\hspace*{5em}{\sc The TCP Algorithm}\vspace{0.5em}\\
 Input: A reticulation-visible network $N$ and a tree $T$, which are binary.\vspace{0.5em}\\
 1. Decompose $N$ into tree node components: $C_t\prec  C_{t-1} \prec  \cdots \prec  C_1$, \\
 ~~~where $\prec $ is a topological order such that that  no directed path \\
~~~ from a node $u$ in $C_j$ to a node $v$ in $ C_i$ exists if $i>j$;\\
 2. $N'\leftarrow N$ and $T'\leftarrow T$;\\
 3. {\bf Repeat} unless ($N'$ becomes a single node) \{\\
 \hspace*{1em} 3.1. Select the lowest big tree node component $C$;\\
 \hspace*{1em} 3.2. Compute $L_C$ in Eqn.~(\ref{eq0}) and select $\ell \in L_C$;\\
 \hspace*{1em} 3.3. Compute the path $P_{T}$ from the root to $\ell$  in Eqn.~(\ref{eq1});\\
\hspace*{1em} 3.4. Determine the smallest index $s_C$ defined by Eqn.~(\ref{eq2});\\
\hspace*{1em} 3.5. Determine the smallest index $d_C$ defined by Eqn.~(\ref{eq3});\\
\hspace*{1em} 3.6. {\bf If} ($s_C > d_C$), output ``$N$ does not display $T$";\\
\hspace*{1em}~~~~~~~{\bf else} \{\\
\hspace*{4em} For each $r\in \mbox{CR}(C)$ $\{$\\
\hspace*{5em} if ($c(r)\not\in \mathcal{D}_T(v_{d_C})$), delete $(z, r)$ for $z\in p(r)\cap {\cal V}(C)$;\\
\hspace*{5em} if ($c(r)\in \mathcal{D}_T(v_{d_C})$), delete $(z, r)$ for $z\in p(r)\backslash {\cal V}(C)$;\\
\hspace*{4em} $\}$\\
\hspace*{4em} Replace $C$ (resp. ${\cal D}_T(v_{d_C})$)  by a leaf $\ell_C$ in $N'$ (resp. $T'$);\\
\hspace*{4em} Remove $C$ from the list of tree node components;\\
\hspace*{4em} Update $\mbox{CR}(C')$  for the affected big tree node components $C'$; \\ 
\hspace*{3em} \} /* {\tt end if} */\\
\hspace*{1em} \} /* {\tt end repeat} */\\
 \bottomrule
\end{tabular}
\end{center}
\vspace*{1em}

We now analyze the time complexity of the { TCP 
Algorithm}. 
  Step 1 can be done in $O(|{\cal E}(N)|)$ time if a breadth-first search is used. 

Step 3 is a while-loop. During each execution of this step, 
the current network is obtained from the previous network by replacing the big tree node component examined in the last execution with a new leaf node. 
Because of this, the modification done in the last two lines in Step 3.6 makes the tree decomposition of the current network  available before the current execution.   Step 3.1 takes a constant time. 
The time spent in Step 3.2 for each execution is $O(|{\cal E}(T_C)|)$.   Step 3.3 takes $O(|{\cal V}(T)|)$ time. 

By Proposition~\ref{Prop_51}, the total  time spent in Step 3.4  is $O(|{\cal L}(N)|^2)$. 
By Proposition~\ref{Prop_55}, the total  time spent in Step 3.5  is 
$\sum_{i} O(|{\cal V}(T_{C_i})||{\cal L}(T)|)$, which is $O(|{\cal E}(N)|\cdot|{\cal L}(T)|)$. 
The time spent in Step 3.6 for each execution is $O(\sum_{u\in {\cal V}(C)} |c(u)|)$. Hence, the total time spent in Step 3.6 is 
$O(|{\cal E}(N)|$.

Taken altogether, these facts imply that  the {TCP Algorithm} takes quadratic time,  
proving the following theorem,

\begin{theorem}\label{main-theorem}
Given a  reticulation-visible network $N$ and a phylogenetic tree with the same labeled leaves as  $N$, the TCP for $N$ and $T$ can be solved in    $O(|{\cal E}(N)|\cdot |{\cal L}(T)|)$ time. 
\end{theorem}

\section{A Linear-time Algorithm for the CCP}
\label{sec6}

As another application of the decomposition theorem, we shall present a  linear-time  algorithm for the CCP in this section.
Given a binary reticulation-visible network $N$ and a subset $B\subseteq {\cal L}(N)$, we would like to determine whether or not  $B$ is a cluster of some node in a tree displayed by $N$.
The edges entering a reticulation node are called {\it reticulation edges} in this section. 

 Assume that $N$ has $t$ big tree node components
 $C_1, C_2, \cdots, C_t.$  
Consider the lowest big tree-node component $C$. 
We use the same notation as in Section~\ref{sec5}: $L_C$ is defined in Eqn.~(\ref{eq0});   $\rho(C)$ denotes  the root of $C$; $\mbox{IR}(C)$ and $\mbox{CR}(C)$ denote  the set of 
inner and cross-reticulations below $C$, respectively. 
Finally, we set 
$\bar{B}={\cal L}(N)\backslash B$.

When $L_C\cap B \neq \emptyset$ and $L_C\cap \bar{B} \neq \emptyset$,    $L_C$ contains two leaves 
$\ell_1$ and $\ell_2 $  such that 
 $\ell_1 \in B$, but $\ell_2 \not\in B$.  
  If $B$ is the cluster of a node $z$ in a subtree $T'$ of $N$,  $z$ is in the unique path $P$
from   $\rho(T')$ ($=\rho(N)$)  to $\ell_1$ in $T'$.  

Assume that $z$ is between $\rho(N)$ and  $\rho(C)$ in $P$.  Since $\rho(C)$ is visible on $\ell_2$, $\ell_2$ is below $\rho(C)$ no matter which incoming edge is contained in $T'$ for  each $r\in \mbox{IR}(C)$. This implies that $\ell_2$ is below $z$ and thus in $B$, which is  a contradiction.  Therefore, if  $B$ is a soft cluster, it must be  a soft cluster of a node in $C$.

When  $L_C \cap \bar{B}=\emptyset$ (i.e., $L_C \subseteq B$),   
we construct a subnetwork $N'$ of $N$ by deleting: \vspace{-0.5em}
\begin{itemize}
   \item  all but one of the incoming edges for each $r\in  \mbox{IR}(C)$,\vspace{-0.5em}
   \item all incoming edges whose tails are in $C$ for  each $r\in \mbox{CR}(C)$ such that $c(r)\not\in B$, and \vspace{-0.5em}
   \item  all incoming edges but one with a tail in $C$ for  each $r\in \mbox{CR}(C)$ such that $c(r)\in B$. \vspace{-0.5em}
\end{itemize}
Note that ${\cal D}_{N'}(\rho(C))$ is a subtree with the following set of network leaves:
\begin{equation}
\hat{B}=L_C \cup  \{ c(r) \;|\; r\in 
\mbox{CR}(C) \mbox{ s.t. } c(r)\in B\}. 
\label{B_def}
\end{equation}
%
Hence, if $B=\hat{B}$, $B$ is then  the  cluster of $\rho(C)$  in $N'$. If
$ \hat{B}\subsetneq B$, we replace ${\cal D}_{N'}(\rho(C))$ by a new leaf $\ell_C$ and  
and set $B'=\{\ell_C\} \cup \left(B\backslash \hat{B}\right)$.  
\vspace{0.5em}

\noindent 
\begin{proposition}
\label{Prop_61}
Assume that $ \hat{B}\subsetneq B$. 
$B$ is a soft cluster in $N$ if and only if $B'$ is a soft cluster in $N'$.
\end{proposition}

\begin{proofproof} 
Recall that 
$B' = (B  \cup \{\ell_C\}) \backslash \hat{B}$. Assume that $B'$ is the cluster of a node $z$ in a tree $T''$ displayed in $N'$.
When $N'$ is reconstructed, $\ell_C$  replaces  the subtree $T'$ rooted at $\rho(C)$ for which the leaves are $\hat{B}$; so if we re-expand $\ell_C$ into $T'$, the cluster of $z$ in $N$ becomes $\left( B \backslash \hat{B}\right)  \cup \hat{B} = B$ and thus $B$ is a soft cluster in $N$.

Assume that $B$ is  the cluster of a node $z$ in a subtree $T$  of $N$.
Let $E$ be the set of reticulation edges removed to obtain $T$ from $N$.
Since $B\neq \hat{B} = B \cap {\cal L}({\cal D}_N(\rho(C)))$, there is a leaf $\bar{\ell}\in B$ that is not below $\rho(C)$. Since  $\bar{\ell}$ is below $z$ in $T$,  $z$ must be above $\rho(C)$ in $T$. 

Consider a reticulation node $r\in \mbox{CR}(C)$ such that $c(r)\in B$.  Since 
$c(r)$ is in $B$, it is  a leaf below $z$ in $T$. 
By our definition of cross-reticulation,  $r$ has at least one parent in $C$. Let $(p_r, r)$ be an edge such that $p_r \in C$.
Note that all but one of the incoming edges of $r$ are in $E$. 
 Define:
\begin{eqnarray*}
  E'& =& 
      \left[ E  \cup  \{ (p, r) \in {\cal E}(N) \;|\; r\in \mbox{CR}(C)  \mbox{ s.t. } c(r)\in B  \; \& \; p\notin {\cal V}(C) \}\right]\\
  && - \{(p_r, r) \;|\; r\in \mbox{CR}(C) \mbox{ s.t. } c(r)\in B\}.
\end{eqnarray*}  
It is not hard to see that  $(p_r, r)$ is the unique incoming edge of $r$ that is not in $E'$ for each $r \in \mbox{CR}(C)$ such that $c(r)\in B$.

 Let $T'= N - E'$. 
$T'$ may contain some dummy leaves that are internal nodes in $N$. 
 It is easy to see that the cluster of $z$ is equal to $B$ and  $\hat{B}$ is the cluster of $\rho(C)$ in $T'$. If we contract the subtree below $\rho(C)$ into a single leaf $\ell_C$,  the cluster of $z$ becomes $B \cup \{\ell_C \} \backslash \hat{B}$, which is $B'$. Therefore,  $B'$ is a soft cluster  in $N'$.
\end{proofproof}

When $B\cap L_C =\emptyset$,  $B$ may or may not be  a soft cluster  in ${\cal D}_N(\rho(C))$. 
Assuming that $B$ is not a soft cluster in ${\cal D}_N(\rho(C))$,  we  reconstruct $N'$ from $N$ by:\vspace{-0.5em}
\begin{itemize}
\item removing all the  edges in $\{(u, r)\in {\cal E}(N) \;|\; r\in \mbox{CR}(C) \mbox{ s.t. } c(r)\notin B, u\not\in  {\cal V}(C) \}$,\vspace{-0.5em}
\item   removing all the edges in $\{(u, r)\in {\cal E}(N) \;|\; r\in \mbox{CR}(C) \mbox{ s.t. } c(r)\in B, u\in {\cal V}(C)\}$, and\vspace{-0.5em}
   \item  replacing ${\cal D}_N(\rho(C))$ by a new leaf $\ell_C$.\vspace*{-0.5em}
\end{itemize}
Similar to the last case, we present the following proposition.

\begin{proposition}
\label{Prop_62}
Assume that $B$ is not in ${\cal D}_N(\rho(C))$ and $L_C \cap B=\emptyset$.  
$B$ is a soft cluster  in $N$  if and only if $B$ is a soft cluster in $N'$.
\end{proposition}

\begin{proofproof}
$N'$ is a subnetwork of $N$. If $B$ is a soft cluster in $N'$,  it is a soft cluster in $N$.

Conversely, assume that $B$ is  the cluster of a node $z$ in a subtree $T$  of $N$.
Let  $E$ be the set of reticulation edges that are removed to obtain $T$ from $N$.
By our assumption,  $B$ is not a soft cluster in ${\cal D}_N(\rho(C))$.  Since $\rho(C)$ is visible on all leaves in $L_C$,   $\rho(C)$ is not below $z$ in $T$.
Therefore, neither $\rho(C)$ nor $z$ is an ancestor of the other in $T$. 

Consider a reticulation $r \in\mbox{CR}(C)$. Since $r$ is a cross-reticulation, it has at least one parent in $C$.
We select a parent $p_r$ in $C$. 
If $c(r)\in B$, $c(r)$ is a leaf below $z$ in $T$ and thus  the unique  incoming edge of $r$  contained  in $T$  is from a node that is not in $C$ to $r$. 
If $c(r)\not\in B$,  the unique incoming edge contained in $T$  may or may not have a tail in $C$.  However, if this edge is not between a node in $C$ and $r$, we can obtain a tree in which $z$ still has $B$ as its cluster by replacing  the edge
 with the edge $(p_r, r)$, where $p_r$ is the selected parent of $r$ in $C$. Therefore, 
we  define:
\begin{eqnarray*}
E'&=&  [E  \cup \{(p, r)\in {\cal E}(N) \;|\; r\in \mbox{CR}(C) \mbox{ s.t. } c(r)\not\in B,  p\in p(r) \}]\\
 && - \{(p_r, r) \;|\; r\in  \mbox{CR}(C) \mbox{ s.t. } c(r)\not\in B \}.
\end{eqnarray*}
Note that $(p_r, r)$ is the unique incoming edge of $r$ that is not in $E'$ for each $r\in  \mbox{CR}(C)$ such that  $ c(r)\not\in B$. 

Let  $T'= N - E'$. $T'$ may contain some dummy leaves that are internal nodes in $N$.  However, it is easy to see that the cluster of $z$ in $T'$ remains the same as the cluster of $z$ in $T$, which is equal to $B$. 
 If we contract $\mathcal{D}_{T'}(\rho(C))$ into a single leaf $\ell_C$, $T'$ is a subtree of $N'$, implying that $B$ is a soft cluster in $N'$. 
\end{proofproof}

We next show how to determine  whether or not $B$ is in $C$ in linear time.
Let $T_C$ be the tree defined in Eqn.~(\ref{eq44}) and (\ref{eq55}). For each $r\in \mbox{IR}(C)$,  $A_{r}$ denotes  the set of ambiguous leaves  defined in Eqn.~(\ref{ar-def}) and $\mbox{lca}_{T_C}(r)$ denotes the LCA of the leaves in $A_{r}$. 



\begin{table}[!b]
\begin{center}
\begin{tabular}{l}
 \toprule
 \hspace*{5em} {\sc Algorithm 1}\\
 Input:  $T_C$ and a subset $B$ of leaves in ${\cal D}_{N}(\rho(C))$. \smallskip\\
 
1. If $|B|==1$, {output} ``Yes" and {\bf exit};\\
2. Construct $T_C$ as defined in Eqn.~(\ref{eq44}) and (\ref{eq55});\\
3. Pre-process $T_C$ so that the LCA of any two  nodes can be found in $O(1)$ time;\\
4. Traverse the leaves in $T_C$ to compute the nodes in $V_{\mbox{lca}}$; \\
5. For each  leaf $\ell \not\in A(C) \cup O(C)$ such that $\ell\not\in B$,\\
~~~~~~~mark the nodes in the path from $\rho(T_C)$ to it;\\
 ~~~For each $r\in \mbox{IR}(C)$ such that $c(r) \not\in B$,\\
~~~~~~~mark the nodes in the path from $\rho(T_C)$ to $\mbox{lca}_{T_C}(r)$ inclusively;\\
6. Traverse the nodes $u$ in $T_C$ to compute the nodes in $V_{\max}$: \\
~~~~~~  check if $u$ is unmarked and its parent is marked in Step 5 when visiting $u$;\\
7.  For each node $u\in V_{\max}$ $\{$\\
~~~~~~ 7.1  Check whether or not all leaves in $B$ are below $u$;\\
~~~~~~ 7.2  Output ``Yes" and {\bf exit} if so; \\
8. Output ``No" and {\bf exit}; \\
\bottomrule
\end{tabular}
\caption{An algorithm to decide whether $B$ is a soft cluster in $C$. }
\label{table4}
\end{center}
\end{table}
 
\begin{proposition}
\label{prop65}
 {\rm  (i)} Let $\ell$ be a leaf in $T_C$ that is  neither ambiguous nor optional.  If $\ell \notin B$,  $B$ is not 
a soft cluster of any node $u$  in the path from $\rho(C)$ to $\ell$ in $N$. 

 {\rm (ii)}  For each $r\in \mbox{IR}(C)$ such that $c(r)\not\in B$,   
$B$ is not a soft cluster of any $u$  in the path from $\rho(C)$ to $\mbox{\rm lca}_{T_C}(r)$ inclusively in $N$. 
\end{proposition}
\begin{proofproof}
This can be proven in the same way as Proposition~\ref{prop565}. \end{proofproof}


Let $\bar{T}_{B}$ be the spanning subtree of $T_C$ over $\{ \ell \in {\cal L}(T_C) \;|\; 
\ell \not\in A(T_C)\cup O(T_C)\cup B\} \cup V_{\mbox{lca}} \cup \{\rho(C)\}$, where $A(T_C)$ and $O(T_C)$  are the sets of ambiguous and optional leaves in 
$T_C$, respectively, and $ V_{\mbox{lca}}$ is defined in Proposition~\ref{prop564}.  
We further define $V_{\max}=\{ v\in {\cal V}(T_C) \;|\; v\not\in {\cal V}(\bar{T}_{B}) \;\mbox{and}\; p(v)\in {\cal V}(\bar{T}_{B})\}$. 

\begin{proposition}
\label{prop66}
$B$ is a soft cluster  in $\cD_{N}(\rho(C))$ if and only if  a node $v\in V_{\max}$ exists  such that for each $\ell\in B$, there is a leaf below $v$ 
with the same label as $\ell$.
\end{proposition}
\begin{proofproof}
Assume that $B$ is a soft cluster of a node $u$ in $\cD_{N}(\rho(C))$. By Proposition~\ref{prop65}, $u$ is not in $\bar{T}_{B}$ and thus  it is below some 
$v\in V_{\max}$. For any $\ell \in B$, $u$  and hence $v$ have  a common leaf descendant with the same label as $\ell$.

Let $v\in V_{\max}$ satisfy the property that for each $\ell\in B$,  a leaf $\ell'\in {\cal L}(T_C)$ exists that has the same label as $\ell$.  
For each $x\in \mbox{IR}(C)$ such that $c(x)\not\in B$, by the definition of $V_{\max}$,  $A_x$ contains an ambiguous leaf that is not below $v$. We select a parent $p'_x$ of $r$ that is not below $v$ in $C$. 

For each $y\in \mbox{IR}(C)$ such that $c(y)  \in B$, we select a parent $p''_y$  below $v$. 

For each $r\in \mbox{CR}(C)$ such that $c(r) \in B$, we select a parent $p_r$ below $v$.

Set:
\begin{eqnarray*}
E&=&\{ (p, r)\in {\cal E}(N) \;|\; p\in {\cal V}(C), r\in A(T_C)\cup O(T_C)\}\\
          & & - \{ (p'_{x}, x) \;|\; x \in \mbox{IR}(C) \mbox{ such that }  c(x)\not\in B\}\\
      & & - \{  (p''_{y}, y) \;|\;  y\in \mbox{IR}(C) \mbox{ such that } c(r)\in B\}\\
  & & - \{  (p_r, r) \;|\;r\in \mbox{CR}(C) \mbox{ such that }  c(r)\in B\}.
\end{eqnarray*}
Therefore, $\cD_{N}(\rho(C))-E$ is a subtree in which $B$ is the cluster of $v$. It is not hard to see that  $\cD_{N}(\rho(C))-E$ can be extended into a subtree of
$N$.
\end{proofproof}

Taken together, the above facts imply that we can use {\sc Algorithm 1} (Table~\ref{table4}) for determining whether a leaf subset is a soft cluster in the lowest big tree-node component or not.
The correctness of {\sc Algorithm 1} follows from Propositions~\ref{prop65} and \ref{prop66}.

Step 1 takes constant time. Step 2 can be done in $O(\sum _{u\in {\cal V}(C)} |c(u)|)$ time. Step 3 takes $O(|{\cal E}(T_C)|)$ time  (see \cite{Tarjan}).
By Proposition~\ref{prop564}, Step 4 can be done in $O(|{\cal E}(T_C)|)$ time. 
In the proof of Proposition~\ref{Prop_55},  Step 5 can be executed in $O(|{\cal E}(T_C)|)$ time. 
Obviously, Step 6 takes $O(|{\cal E}(T_C)|)$ time.  
For each node $u$,  Step 7.1 takes $O(|{\cal E}(\cD_{T_C}(u))|)$ time. Since  all the examined subtrees
are disjoint, the total time taken by Step 7.1 is  $O(|{\cal E}(T_C)|)$ time.

Taking all the above facts together, we are able to give a linear-time algorithm for the CCP.
\vspace*{0.5em}

\noindent 
\begin{center}
\begin{tabular}{l}
 \toprule
 \hspace*{5em} {\sc The CCP Algorithm}\\
 {\bf Input:} A binary network $N$ and a subset $B \subseteq {\cal L}(N)$. \vspace*{0.5em} \\
1.  Compute the big tree node components sorted in a topological order: \\
~~~~~~~~$C_t \prec  C_{t-1} \prec \cdots \prec  C_1$\\
~~~~~~~~ such that $C_i$ is below $C_j$ only if $i<j$;\\
2.  {\bf for} $k=1$ {\bf to}  $t$ {\bf do} $\{$\\
\hspace*{1em} 2.1.~ Set $C=C_k$;  compute $L:=L_{C_k}$ as defined in Eqn.~(\ref{eq0});\\
 \hspace*{1em} 2.2.~ $Y := \mbox{ (~Is $B$ a soft cluster in ${\cal D}_{N}(\rho(C))$?~)}$;\\
 \hspace*{1em} 2.3.~ {\bf if} ($Y==1$) {output} ``Yes" and {\bf exit};\\
\hspace*{1em} 2.4.~ {\bf if} ($Y==0$) $\{$\\
\hspace*{3.5em}~~~ $\bar{B}:= {\cal L}(N)\backslash B$;\\
\hspace*{3.5em}~~~ {\bf if} ($L\cap \bar{B} \neq \emptyset$ \& $B\cap L \neq \emptyset$) {output} ``No" and {\bf exit};\\
\hspace*{3.5em}~~~ {\bf if} ($B\cap L==\emptyset$) $\{$\\
\hspace*{4.5em}~~~   Remove edges in $\{(u, r) \;|\; r\in \mbox{CR}(C) \mbox{ s.t. } c(r)\not\in B, u\not\in C\}$;\\
\hspace*{4.5em}~~~  Remove edges in $\{(u, r) \;|\; r\in \mbox{CR}(C) \mbox{ s.t. } c(r)\in B, u\in C\}$;\\
\hspace*{3.5em}~~~ $\}$\\
 \hspace*{3.5em}~~~ {\bf if} ($ \bar{B} \cap L==\emptyset$) $\{$\\
\hspace*{4.5em}~~~   Remove edges in $\{(u, r) \;|\; r\in \mbox{CR}(C) \mbox{ s.t. } c(r)\not\in B, u\in C\}$;\\
\hspace*{4.5em}~~~   Remove edges in $\{(u, r) \;|\; r\in \mbox{CR}(C) \mbox{ s.t. } c(r)\in B, u\not\in C\}$;\\
\hspace*{4.5em}~~~  $B:=  \left(B \cup \{\ell_{C} \}\right) \backslash \left(L \cup \{ c(r) \;|\; r\in \mbox{CR}(C) \mbox{ s.t. } c(r)\in B\}\right)$;\\
\hspace*{3.5em}~~~ $\}$\\
\hspace*{3.5em}~~~  Replace ${\cal D}_N(\rho(C))$ by a leaf $\ell_{C}$;\\
\hspace*{3.5em}~~~ Remove $C$ from the list of big tree node components;\\
\hspace*{3.5em}~~~ Update $\mbox{CR}(C')$  for the affected big tree node components $C'$; \\ 
\hspace* {1em} ~~~$\}$\\
$\}$ /* for */\\
\bottomrule
\end{tabular}
\end{center}

The CCP algorithm runs in linear time.   Step 1 takes $O(|{\cal E}(N)|)$ time. 
Step 2 is a for-loop that runs $t$ times. 
Since the total number of network leaves in $C_k$ and the reticulation nodes below $C_k$ is $|{\cal E}(C_k)|$ at most, 
Step 2.1 takes $O(|{\cal E}(C_k)|)$ time for each execution. 
In Step 2.2, the linear-time {\sc Algorithm 1}  is called to  compute  $Y$  in  
$O(|{\cal E}(C_k)|)$ time. 
Obviously, Step 2.3 takes constant time.
To implement Step 2.4  in linear time, we need to use an array $A$ to indicate whether a network leaf is in $B$ or not.
$A$ can be constructed in $O(|{\cal L}(N)|)$ time. With $A$, 
each conditional clause in Step 2.4 can be determined in $|L|$ time, which is
$O(|{\cal E}(C_k)|)$ at most. 
Since the total number of inner and cross reticulations is $|{\cal E}(C_k)|$ at most, each line of Step 2.4 takes $O(|{\cal E}(C_k)|)$ time at most.   Hence,  Step 2.4 still takes $O(|{\cal E}(C_k)|)$ time. 
Taking all these together, the total time taken by Step 2 is 
$\sum_{1\leq k\leq t} O(|{\cal E}(C_k)|)
= O(|{\cal E}(N)|).$
Therefore,  the following theorem has been  proven. 

\begin{theorem}\label{main-theorem2}
Given a reticulation-visible network $N$ and an arbitrary  subset $B$ of labeled leaves in $N$, the CCP for $N$ and $B$ can be solved in    $O(|{\cal E}(N)|)$ time. 
\end{theorem}

\section{Concluding Remarks}
\label{sec8}

We have presented polynomial-time  TCP and CCP algorithms for arbitrary reticulation-visible networks.
 They rely on the decomposition theorem proven in Section~\ref{sec4}. 

In \cite{Gunawan_2015}, we proved for the first time that the number of reticulation nodes in a 
binary reticulation-visible network is bounded from above by a linear function in the number of the leaves. 
The bound was established using the fact that reticulation-visible networks are tree-based \cite{Steel_2015}, that is, they can be obtained from a tree with the same set of labeled leaves by the addition of some edges between the tree edges.  Using the same technique, we later proved that a binary galled network with $n$ leaves has at most $2(n-1)$ reticulation nodes \cite{Gunawan_20152}. 
  In the present paper, we use the decomposition theorem to derive the same bound for arbitrary galled networks, in which tree nodes are not necessarily binary. Therefore,  we present a new technique for establishing the size of a  network with visibility property.

One  
interesting problem for future research  is  how to extend our study  into fast heuristic TCP and CCP algorithms for arbitrary networks.
Other problems include (a) testing whether two reticulation-visible networks display the same set of binary trees in polynomial time and (b) application of the decomposition theorem in reconstructing reticulation-visible networks from gene trees or sequences.   Solutions for these questions are definitely valuable in phylogenetics.

\section*{Acknowledgments} 
The authors are grateful to Philippe 
Gambette, Anthony   Labarre, and St\'ephane   Vialette	for discussions on the problems studied in this work. 
This work was supported by a Singapore MOE ARF Tier-1 grant R-146-000-177-112 and the Merlion Programme 2013.
DasGupta was supported by NSF grant IIS-1160995

\end{document}